\documentclass[11pt]{article}
\usepackage{axodraw}
\usepackage{amsfonts} \parindent 0pt \parskip.2cm \topmargin -1.0cm
\oddsidemargin=0.25cm\evensidemargin=0.25cm \newfont{\bbbold}{msbm10
  scaled \magstep1} \def\com{\mbox{\bbbold C}}

 \newcommand{\intk}{\int \! d^4 k \;}

\newcommand{\xz}{\times} \newcommand{\del}{\partial}
 \newcommand{\f}{\phi} 
\def\th{{\theta}} \newcommand{\fp}{\frac1{(2\pi)^{\frac32}}}
 
 \def\a{\alpha} 
\def\b{\beta}  \newcommand{\adt}{\dot\alpha}
\newcommand{\bdt}{\dot\beta}  
\def\t{\tau} \newcommand{\D}{{\Delta}} 
 \newcommand{\m}{{\mu}} \newcommand{\n}{{\nu}}
 
\newcommand{\thb}{{\mbox{\boldmath{$\theta$}}}}
\def\tfrac#1#2{\begin{array}{c}\hspace{-5pt}\frac{#1}{#2}\hspace{-5pt}\end{array}}
\def\cO{{\cal O}}

 \def\eqs#1#2{(\ref{#1},\ref{#2})}

\def\pslash#1{{\setbox0=\hbox{$#1$}
    \rlap{\ifdim\wd0>.7em\kern.22\wd0\else\kern.1\wd0\fi /}#1}}

\def\be{\begin{equation}} \def\ee{\end{equation}}
\def\ba{\begin{array}} \def\ea{\end{array}}
 \newcommand{\eq}[1]{(\ref{#1})}

\begin{document}
\begin{flushright}
LU-ITP 2004/042 \\
 DAMTP-2004-124
\end{flushright}

\begin{center}
{\Large\sf Quantized equations of motion in non-commutative theories }\\[10pt]
\vspace{.5 cm}

{Paul Heslop${}^1$, Klaus Sibold${}^2$}
\vspace{1.0ex}

\end{center}

{\small \begin{enumerate} \item  Department of Applied Mathematics and
 Theoretical Physics,
 Centre of Mathematical Sciences, Cambridge
University, Wilberforce Road, Cambridge CB3 0WA, UK \\ 
\item Institut f\"ur Theoretische Physik, Universit\"at Leipzig,
Augustusplatz 10/11, D-04109 Leipzig, Germany
\end{enumerate}}

\vspace{2.0ex}
\begin{center}
{\bf Abstract}
\end{center}

\noindent
Quantum field theories based on interactions which contain the Moyal star product
suffer, in the general case when time does not commute with space, from several
diseases: quantum equation of motions contain unusual terms, conserved currents can not
be defined and the residual spacetime symmetry is not maintained. All these 
problems have the same origin: time ordering does not commute with taking the
star product. Here we show that these difficulties can be circumvented by a new
definition of time ordering: namely with respect to a light-cone variable. In
particular the original spacetime symmetries  $SO(1,1)\xz SO(2)$ and translation
invariance turn out to be respected. Unitarity is guaranteed as well.

\newpage
\section{Introduction}
Space and time will, at extremely short distances, require new notions in both 
mathematical description and physical content. A simple physical argument for this
is based on the uncertainty principle which says that black holes can be formed
thus leading to a horizon and other consequences when precision in time is high enough~\cite{DoplicherFredenhagenRoberts}. As a modest step into this direction
one may understand the introduction of Moyal products in otherwise rather
conventional flat space-time quantum field theory. They arise when the coordinates
are being considered as Hermitian operators which satisfy simple commutation 
relations like
\begin{equation}
\left[x_\mu, x_\nu\right] = i\theta_{\mu\nu}.
\end{equation}
A typical interaction reads then
\begin{equation}
  S_{int}=g\int d^4x \,\f(x)*\f(x)*\f(x).
\end{equation}
The discussion for the case when time/space commutators vanish is fairly advanced,
whereas the case when they do not vanish is not yet very well understood.
Although Feynman rules have been proposed which lead to unitarity in non-gauge
theories\cite{LiaoSiboldI, LiaoSiboldII, korean, Bahns:2002vm}, gauge theories seem to be
inconsistent \cite{Liaounpublished, W"urzburg}. A somewhat more detailed
study also reveals that quantum equations of motion have a form which is
intractable in practice \cite{Reichenbachdiplomathesis}, but worse is the
fact that symmetries which are present on the classical level do not seem
to be maintained after quantization.  The main truly disturbing example is
$SO(1,1)\xz SO(2)$ invariance \cite{Reichenbachdiplomathesis}.This is
of course not tolerable: we wish to characterise theories by their symmetry
content, hence every deviation from a classically realised symmetry must be
very well understood until we accept it as unavoidable.\\

In the present paper we first recall the symmetry content on the classical level
as being generically $SO(1,1)\xz SO(2)$. Since the conventional time ordering is not in accordance
with this symmetry and thus the reason for its breakdown, we define a
new notion of time ordering and explore the consequences of this change. It
turns out that the perturbation theory formulated on this basis has all desired
properties: it is compatible with the symmetry, leads to simple Feynman rules
and closed expressions for the quantum equations of motion. The LSZ asymptotic condition
can be formulated and unitarity is maintained. All of this will be derived for
scalar field theories as example and is still restricted to the tree approximation
(apart from the unitarity relations which involve one-loop contributions).
But the generality of the results supports the hope that proceeding into direction of
renormalisation and incorporating gauge theories will be possible.\\

\section{Symmetries and standard form}

We would like to show first that all non-commutative field theories
defined from an action via the Moyal
product are either $SO(1,1)\xz SO(2)$ invariant, or have the symmetries
of the so-called light-like case (the product of two null rotations).
To see this consider for simplicity a non-commutative scalar field
theory (for example scalar $\phi_*^3$ theory.)  The action
$S[\phi;\theta]$, is a functional of the fields $\phi$ and 
a function of the constant $\theta$ matrix $\theta^{\m \n}$. Now given
such an action with an arbitrary $\theta$ we
change basis so that a point previously specified by coordinates $x^{\mu}$
is now specified by coordinates $x'^{\m}=L^{\m}{}_{\n}x^{\n}$ where
$L\in SO(1,3)$. Then the scalar
field $\phi$ transforms to $\f'$ defined as $\f'(x)=\f(L^{-1}x)$ and
it follows that derivatives of the scalar field $\del_{\mu}\phi$ transforms to
$L^{\n}{}_{\m}\del_{\n}\f'$. One
can then see for a theory whose Lorentz violation comes only from the
Moyal star, we have
\begin{equation}
S[\f;\theta]=S[\f';L\theta L^{\rm{T}}].
\end{equation}
So by a simple change of coordinates (ie we make no physical change)
the transformed action has a similar form to the original action but
with $\theta$ replaced by $L\theta L^{\rm{T}}$. This means that
starting with any theory defined in terms of an arbitrary $\theta$ we
may change coordinate basis in order that $\theta$ has the simplest
form possible. 

Let us note parenthetically that the situation has a strong analogy with a broken internal
symmetry. The $\theta$ can be thought of as a field taking on a
constant expectation value which then does not propagate. In this case
the fields arrange themselves in representations of the larger
symmetry although in fact only the symmetry which leaves the
expectation value invariant is preserved. Similarly here we expect to
be able to use the field representations of the full Lorentz symmetry
even though this has been broken down to a smaller
group. 

In order to find the simplest form for $\theta$ it is easier to work
with the spinorial representation of $\theta$. We have
\begin{equation}
\theta^{\m\n}=\t^{\m\n}_{\a\b}{\thb}^{\a\b}+ \overline\t^{\m\n}_{\adt\bdt}\overline
{\thb}^{\adt\bdt}
\end{equation}
where $\a,\b=(1,2)$ are Weyl spinor indices which transform under
$SL(2;\com)\sim SO(1,3)$ and $\tau$ are the Pauli matrices. So
$\theta^{\m\n}$ is equivalent to a complex symmetric $2\xz2$ matrix
$\thb^{\a\b}$ transforming as $\thb'=M\thb M^{\rm{T}}$ where $M\in
SL(2;\com)$ is a complex $2\xz2$ matrix with unit determinant. 
It is easy to check that as long as
the determinant of $\thb$ is nonzero there exists an $M\in SL(2;\com)$
such that
\begin{equation}
\th'=M\th M^{\rm{T}} =\sqrt{\rm{det}(\thb)}I_2. 
\end{equation}
The remaining symmetry which leaves $\th'$ invariant is clearly
$SO(2;\com)$.  

If, on the other hand, the determinant of $\thb$
vanishes then either 
$\thb=0$ or there is an $M$ such that
\begin{equation}
 \th'=M\th M^{\rm{T}} =\left(\ba{cc}1&0\\0&0\ea\right). 
\end{equation}
This corresponds
to the `light-like' case of Aharony, Gomis and
Mehen~\cite{Aharony:2000gz}. 
The remaining
symmetry in this case is given by $2\xz2$ matrices of the form
\begin{equation}
M=\pm \left(\ba{cc}1&b\\0&1\ea\right) \qquad b\in\com.
\end{equation}

If one then translates this back we find that $\th^{\m\n}$ is always
equivalent to one of the following forms
\begin{equation}
  \label{canth}
  \left(\ba{cccc} 0&\th_e&0&0\\
    -\th_e&0&0&0\\0&0&0&\th_m\\0&0&-\th_m&0
   \ea\right) \qquad
 \left(\ba{cccc} 0&0&0&-1\\0&0&0&-1\\0&0&0&0\\1&1&0&0
   \ea\right).
\end{equation}
In the first case the remaining symmetry is $SO(1,1)\xz SO(2)$ in
general, extended to $O(1,1)\xz SO(2)$ if $\th_e=0,\th_m\neq0$, to
$SO(1,1)\xz O(2)$ if $\th_e\neq0,\th_m=0$ and of course to the full
$SO(1,3)$ if $\th_e=\th_m=0$. In the latter case, the `light-like
case' the remaining symmetry is harder to describe. It consists of two
`null rotations' (see~\cite{Simon:2002ma}) both of which leave $x^0-x^1$
invariant. The symmetry is given by $x^{\m}\rightarrow
L^{\m}{}_{\n}x^{\n}$ with
\begin{equation}
L^{\m}{}_{\n} = \left(\ba{cccc} 1+\frac12(a^2+b^2) &
  -\frac12(a^2+b^2)&a&b\\
\frac12(a^2+b^2) &
  1-\frac12(a^2+b^2)&a&b\\
a&-a&1&0\\
b&-b&0&1
\ea\right)
\end{equation}

Since for {\em any} $\theta$ we can choose coordinates such that in
the new coordinates $\theta$ takes one of the forms of
equation~\eq{canth}, it follows that {\em any} theory defined by a
non-zero $\theta$ is invariant under either $(S)O(1,1)\xz(S)O(2)$ or
the afore-mentioned symmetry of the light-like case. Of course in the
original coordinates then these symmetries will, in general, be
difficult to see.

Furthermore we see that in the space of non-commutative theories
almost all cases can be given in terms of a $\theta$ of the
form~(\ref{canth}a) with $\th_e\neq0, \th_m\neq0$. This is thus the
generic case and the case which we will concentrate on in the rest of
this paper.

\section{Locality properties and time ordering}

\subsection{Commutation relations}

It is obvious from the definition of the Moyal product that
the locality properties of the theory will drastically
differ from those of an ordinary quantum field theory. 
To begin with let us consider commutators of composite operators in 
an ordinary  free theory.
We have real scalar fields $\Phi(x)$ which we split into positive and
negative frequency parts as $\Phi(x)=\Phi^+(x)+\Phi^-(x)$ in the usual
way. We canonically quantize the theory and define the commutator
functions
\begin{eqnarray}
  i\Delta^+(x-y)&=&\left[ \Phi^+(x),\Phi^-(y)\right]\\
  i\Delta^-(x-y)&=&\left[ \Phi^-(x),\Phi^+(y)\right]=-i\Delta^+(y-x)\\
  i\Delta(x-y)&=&\left[
    \Phi(x),\Phi(y)\right]=i\Delta^+(x-y)+i\Delta^-(x-y)
\end{eqnarray}
Using standard identities of commutators one finds
\begin{eqnarray}
\left[\Phi(x),\Phi^2(y)\right]&=&2[\Phi(x),\Phi(y)]\Phi(y)=2i\D(x-y)\Phi(y)\\
\left[\Phi^2(x),\Phi^2(y)\right]&=&2i\D(x-y)\left(\Phi(x)\Phi(y)+\Phi(y)\Phi(x)
 \right)
\end{eqnarray}
Both of which are proportional to $\D(x-y)$ and thus have 
support within the light-cone.
Indeed the 
commutator of any two (Wick ordered) monomials of the fundamental fields and a finite
number of derivatives can be written as a sum of terms proportional to
$\D(x-y)$ and derivatives thereof, and so these will also have
light-cone support. 
So operators formed by monomials and a finite number of derivatives
commute at space-like distances. (It is well known that the converse
is also true \cite{Epstein,Schroer}.)

Problems occur however if one considers monomials containing an
infinite number of derivatives such as $\Phi * \Phi(x)$. For example
\begin{eqnarray}
  \label{com*}
  \left[\Phi(x),\Phi*\Phi(y)  \right]&=&i\D(x-y)*_y
  \Phi(y)+i\Phi(y)*_y \D(x-y)
\end{eqnarray}
and the presence of the star can spoil the support properties of the
commutator. This commutator consists of four terms similar to 
\begin{equation}
  \ba{rcl}
  &&i\D^+(x-y)*_y
  \Phi^+(y)+i\Phi^+(y)*_y \D^+(x-y)\\
  &=&\frac1{(2\pi)^{9/2}} \int \frac{d^3k d^3k'}{4k^0k'{}^0}
e^{-ik^{+}y}A(\vec k)e^{-ik'{}^+(x-y)}\left( e^{-i k'^+\wedge k^+}+e^{-i
      k^+\wedge k'^+}\right)\\
  &=&\frac1{(2\pi)^{3/2}} \int
\frac{d^3k}{2k^0}
e^{-ik^+y}A(\vec k)\left(\D^+(x-y+\tilde k/2)+\D^+(x-y-\tilde k/2)\right)
\ea
\end{equation}
where
\begin{equation}
  \label{eq:def}
  *_y=e^{\frac i2\theta^{\m\n} \overleftarrow{\del}_{\m}
    \overrightarrow{\del}_{\n}}\qquad
  k\wedge k'=\frac12 k^{\m}\th_{\m\n}k'^{\n}\qquad
  \tilde k^{\m}=\th^{\m\n}k_{\n}^+\qquad k^+=(\omega_k,\vec k).
\end{equation} 
The commutator is no longer proportional to $\D(x-y)$ but is shifted
by an amount depending on $k$ which is integrated over. Thus in
general the commutator no longer has support only within the
light-cone, and the operators do not commute at space-like
distances. In general equal time commutators will not vanish.

Note however that in the special case $\th_e=0$ then $\tilde
k^0=\tilde k^1=0$ and so the shift only occurs in the $x^2,x^3$
direction. In this case equal time commutators still vanish 
and the support properties become `wedge-like' (see for
example~\cite{Alvarez}).  
Note that it is also possible to define the free theory so that this also
has only wedge-like support properties and only has the symmetry
$SO(1,1)\times SO(2)$ but not the full four-dimensional Lorentz group
(see~\cite{HS2}).

\subsection{Interaction - tree approximation: symmetry breaking} 

We define time ordered Green functions via the Gell-Mann Low
formula:
\begin{equation}\label{gml}
\left\langle T\Phi(x_1)\dots\Phi(x_n)\right\rangle=\left\langle
  T\Phi(x_1)\dots\Phi(x_n)e^{iS_{\rm{int}}}\right\rangle_0
\end{equation}
where $S_{int}$ is the interaction part of the action. On the left
hand side of this equation we have interacting fields and the
expectation value is with respect to the interacting vacuum whereas on
the right-hand side we take free fields and the free vacuum which we
indicate by the subscript~$0$.

Usually the time ordered product of two operators $O_1,O_2$ is defined 
as
\begin{equation}
  \label{eq:T}
  T O_1(x)O_2(y)= \theta(x^0-y^0) O_1(x)O_2(y) +  \theta(x^0-y^0)
    O_2(y)O_1(x).
\end{equation}
Now $\theta(x^0-y^0)$ is not 
 Lorentz invariant.  However the time ordered product defined by \eq{eq:T} {\it is}
 Lorentz invariant provided $O_1(x)$ and $O_2(y)$ commute at
 space-like distances. 
 This is because if $x-y$ is time-like then $\theta(x^0-y^0)$ is
 Lorentz invariant 
 whereas if $x-y$ is space-like then the order is irrelevant
and 
\begin{equation}
TO(x)O(y)=(\theta(x^0-y^0)+  \theta(x^0-y^0)) O_1(x)O_2(y)=O_1(x)O_2(y)  
\end{equation}
which is also Lorentz invariant.

This is no longer true in a general non-commutative field theory
since, as we saw in the previous section, equal time commutators 
no longer vanish (unless we consider the special case with $\theta_e=0$).
So if we use the
above definition of time ordering we do not expect the time-ordered
products to 
obey even the remaining $SO(1,1)\xz SO(2)$ invariance.  We therefore
introduce a new definition of time ordering.

\subsection{Light-wedge variables and new time ordering} \label{lwnto}

In the present case of $SO(1,1)\xz SO(2)$ invariance a suitably
adapted time ordering will be defined in the next subsection hence we
introduce  the respective variables as
\begin{equation}\label{lwv}
u=(x^0-x^1)/\sqrt2,\ v=(x^0+x^1)/\sqrt2.   
\end{equation}
It is useful to re-express the free fields in terms of these variables which
we call `light-wedge variables'. Note that such co-ordinates are used in
light-cone quantization of field theories. We
define momenta (with indices downstairs) as
\begin{equation}
k_u=\frac{k_0-k_1}{\sqrt2},\ k_v=\frac{k_0+k_1}{\sqrt2}   
\end{equation}
and this
means $k^u=k_v,\ k^v=k_u$. So the solution of the Klein Gordon
equation (this is completely equivalent to the usual solution via a
change of variables) becomes
\begin{eqnarray}
  \label{eq:uTf}
  \Phi(x)&=&\fp \intk
\delta(2k_uk_v-k_{a}k_{a}-m^2)A(k)e^{-ikx}\\
&=&\fp \int \frac{d^3k}{2k_v} A({\bf k})e^{-i\overline kx}\label{FA}
\end{eqnarray}
where in the second line $d^3k:=dk_vdk_2dk_3$ and ${\bf
  k}=(k_v,k_{a})$ and the on-shell momentum $\overline k$ 
defined as
\begin{equation}\label{osu}
\overline k_u=(m^2 +k_{a}k_{a})/(2k_v)\quad \overline k_v=k_v \quad \overline k_a=k_a
\qquad a=2,3
\end{equation}
Note that with these variables there is no need to
separate positive and negative frequency parts. Taking $k_v$
positive corresponds to positive frequency and vice versa. The reality
of $\Phi$ implies
\begin{equation}
  \label{eq:real}
   A^{\dagger}({\bf k})=-A(-{\bf k}).
\end{equation}

Inverting~\eq{FA} we can express $A({\bf k})$ in terms of the field
$\Phi(x)$
\begin{equation}\label{invert}
  A({\bf k})=\frac 1{2\pi^{3/2}}\int d^3x 2k_v e^{i\overline kx} \Phi(x).
\end{equation}

We quantize the fields, with the commutation relation
\begin{equation}
  \label{eq:comu}
  \left[A({\bf k}),A({\bf k'})\right]=2k_v\delta^3({\bf k+k'})
  \end{equation}
and the vacuum satisfies
\begin{equation}
  A({\bf k})|0\rangle= 0\quad  k_v<0 \qquad  \langle0|A({\bf k})=
  0\quad k_v>0
\end{equation}

The commutator function has the form
\begin{equation}
  \label{eq:comu2}
  i\D(x-x')=\left[\Phi(x),\Phi(x')\right]=\frac1{(2\pi)^3} 
  \int \frac{d^3k}{2k_v} e^{-i\overline{k}(x-x')}
\end{equation}
and $\D^+$ ($\D^{-}$) are given by similar expressions but with the
$k_v$ integration restricted to the interval $(0,\infty)$ ($(-\infty,0)$).
One can explicitly check using Bessel function identities such as (in
the two dimensional case)
\begin{equation}
  \label{eq:Bessel}
  (\pi i/2) H_0^{(1)}(-2m\sqrt{uv})=\int_0^{\infty}\frac{d
  k_v}{2k_v}e^{-i m/k_u (u)+k_v(v)}
  \end{equation}
  that these give the same expressions as in the usual case.
  
  Finally, we will wish to redefine the causal Green function according
  to the new time ordering as
\begin{equation}
  \label{eq:Dcu}
  \D_c(x)=\th(u)\D^+(x)-\th(-u)\D^-(x).
\end{equation}
In fact this is identical to the standard propagator defined in
terms of the usual time ordering as we argue below.

\subsection{$SO(1,1)\times SO(2)$ invariant time ordering}

In order to keep
the remaining $SO(1,1)\times SO(2)$ symmetry we use the following definition of
time ordering
\begin{equation}
  \label{eq:T2}
  T O_1(x_1)O_2(x_2)= \theta(u_1-u_2) O_1(x_1)O_2(x_2) +  \theta(u_2-u_1)
    O_2(x_2)O_1(x_1).
\end{equation}
where we have used the `light-wedge' coordinates $u,v$ defined
in~\eq{lwv}.

Note that for two space-like commuting operators this time-ordering is in fact
equivalent to the usual one since for time-like $x$, we have that  
$u>0 \Leftrightarrow x^0>0$.  So we are using  a
choice of time-ordering which is equivalent to the usual prescription for
ordinary theories, but which also maintains the $SO(1,1)\xz SO(2)$
symmetry in the non-commutative case. This is one way of seeing
that the free propagator defined with the $u$ time ordering is
equivalent to the usual time ordering, since the propagator used in ordinary
perturbation theory is just the vacuum expectation value of the time
ordered product of two fundamental fields which are free and do
indeed commute at space-like distances.

To see that this new time ordering respects the symmetry note that
under a  $SO^+(1,1)$ transformation $u\rightarrow a u,\ v\rightarrow
v/a;a>0$ and so $\theta(u)$ is invariant without the  need for
space-like commutativity.

\section{The quantum equation of motion}

\subsection{Usual time ordering}

We wish to consider the tree level quantum equations of motion for a
non-commutative field theory. Eventually we will consider an
interaction term $\f_*^3$ but to illustrate the techniques we first
consider some simple cases. In this subsection we define time-ordering
in the standard way with respect to $x^0$ whereas in the next
subsection we will
use the new  time-ordering with respect to $u$. Firstly consider the
standard case of a 
theory defined by a free Lagrangian together with an interaction
Lagrangian which contains no time derivatives.  We wish to find the
quantum equation of motion for such a theory, ie the
equation of motion for a field inserted into a Green's function
defined using the Gell-Mann Low formula
\begin{eqnarray}\label{line1}
  (\square_x + m^2)\left\langle T\f(x) X\right\rangle &=&(\square_x
  + m^2)\left\langle T\f(x) X
  e^{iS_{\rm{int}}}\right\rangle_0\\
  &=&(\square_x + m^2)\int d^4y\left\langle T\f(x)
  \f(y)\right\rangle_0 \left\langle
  T\tfrac{\delta}{\delta\f(y)}( 
  Xe^{iS_{\rm{int}}})\right\rangle_0\label{line2}\\
  &=& \left\langle
  T\tfrac{\delta S_{int}}{\delta\f(x)} 
  Xe^{iS_{\rm{int}}}\right\rangle_0 + \rm{c.t.}\\
  &=&\left\langle
  T\tfrac{\delta S_{int}}{\delta\f(x)} 
  X\right\rangle + \rm{c.t.}\label{line4}
\end{eqnarray}
Here $X$ represents any monomial of fields and derivatives thereof.
The second line is obtained using Wick's theorem and we will discuss
this further below: it is only valid as written when there are no time
derivatives in $S_{int}$.  In the third line we have used that
\begin{equation}
(\square_x + m^2)\left\langle T(\f(x)
  \f(y))\right\rangle_0 =
-i\delta(x-y) 
\end{equation}

and `c.t.' stands for `contact terms' which arise from
$\frac{\delta}{\delta\f(y)} X$.  Finally we re-express the answer in
terms of interacting fields to obtain the fourth line.  We find a
quantum equation similar to the classical equation up to contact
terms.

As mentioned~\eq{line2} can be derived from Wick's theorem but only
when $S_{int}$ contains no time derivatives: these interfere with the
time ordering. To see this consider for illustration $S_{int}$ of the form
\begin{equation}\label{ill}
S_{int}=g\int d^4x \,\cO\, (\del_0)^n\f(x) 
\end{equation}
where $\cO$ is a monomial in $\phi$. Then using
\begin{equation}
(\square_x +m^2)\left\langle T\left( \f(x)(\del_0)^n\f(y)
\right) \right\rangle_0 =
\left\{ \ba{ll}
-i(\vec{\del_x}^2
-m^2)^{\frac n2}\delta(x-y) \qquad &n \mbox{ even}\\
i\del_0(\vec{\del_x}^2 -m^2)^{\frac{n-1}2}\delta(x-y)
\qquad &n \mbox{ odd} \ea \right.\label{using}
\end{equation}
we find that the quantum equation of motion for $n$ even is
\begin{eqnarray}
&&  (\square_x + m^2)\left\langle T(\f(x) X)\right\rangle \\\label{qe0}
&=&(\square_x + m^2)\ \!i \int d^4y\left\langle T \f(x)
  (\del_0)^n\f(y)\right\rangle_0 \left\langle
  T\cO 
  Xe^{iS_{\rm{int}}}\right\rangle_0\\\nonumber
&+& (\square_x + m^2)\ i \!\int d^4y\left\langle T\f(x)
  \f(y)\right\rangle_0 \left\langle
  T\tfrac{\del \cO}{\del\f}(y)(\del_0)^n\f(y) X  
  e^{iS_{\rm{int}}}\right\rangle_0
\\&=  &g{(\vec{\del_x}^2 \!\!-\!m^2)^{\frac n2}}
\left\langle
  T\left(\cO Xe^{iS_{\rm{int}}}\right)\right\rangle_0 + g\left\langle
  T\left(\frac{\del \cO}{\del\phi} (\del_0)^n\f(x)
  Xe^{iS_{\rm{int}}}\right)\right\rangle_0+\rm{c.t.}\label{qe1}\\
  &=&
g\left\langle
  T\left((\vec{\del_x}^2 \!\!-\!m^2)^{\frac n2}
    \cO(x) Xe^{iS_{\rm{int}}}\right)\right\rangle_0 + g\left\langle 
  T\left(\frac{\del \cO}{\del\phi}(x)(\vec{\del_x}^2
        \!\!-\!m^2)^{\frac 
          n2}\f(x)
  Xe^{iS_{\rm{int}}}\right)\right\rangle_0+\rm{c.t.} \label{qe2}\\\label{qe3}
&=&\left\langle
  T\left(\tfrac{\delta \tilde S_{int}}{\delta\f(x)} 
  X\right)\right\rangle + \rm{c.t.}
\end{eqnarray}
where we define a modified effective interaction as
\begin{equation}\label{mei}
\tilde S_{int} = g\int d^4x \cO (\vec{\del_x}^2
    \!\!-\!m^2)^{\frac n2}  \f(x) \quad \mbox{$n$ even}.
\end{equation}
Notice that if $\f$ is a free field $S(\f)=\tilde S(\f)$ but for a
general field the two actions are different. Thus the manipulations
from~\eq{qe0} 
to~\eq{qe2} work because there we are dealing with free fields, as indicated by
the subscript 0 for the correlators (see~\eq{gml}). But the result for
interacting fields of~\eq{qe3} is non-trivial. 

Some comments on the manipulations above. Equation~\eq{qe0} is obtained via the Gell-Mann Low formula using
Wick's theorem (and  is the analogue of equation~\eq{line2}). To
obtain~\eq{qe1} we have used~(\ref{using}$a$) and integrated out the
delta function. On going
from~\eq{qe1} to~\eq{qe2}, in the first term we 
have moved the differential operator inside the correlation function
(which is allowed since there are no time derivatives) and in the
second term we have used the equations of motion for the free field
sitting in the propagator. 

A crucial point is that we take all derivatives occurring in
the interaction Lagrangian to act {\em before} the time
ordering whereas the integration itself is taken after the time
ordering. 
It is also possible to define a time ordering $T_*$ where all
derivatives occur outside the time ordering and this definition gives
the na\"ive Feynman rules. For a
standard quantum field theory these two definitions differ by local 
terms only and are therefore equivalent after a finite renormalisation 
whereas for a non-commutative field theory the equivalence is not to be
expected.

The case with  $n$ odd is more complicated. In this case
the  quantum equation of motion is 
\begin{eqnarray}
&&  (\square_x + m^2)\left\langle T\f(x) X\right\rangle \\
&=  &-g\, \del_0 (\vec{\del_x}^2 \!\!-\!m^2)^{\frac {n-1}2}
\left\langle 
  T\cO Xe^{iS_{\rm{int}}}\right\rangle_0 + g\left\langle
  T \frac{\del \cO}{\del\phi} (\del_0)^{n}\f(x)
  Xe^{iS_{\rm{int}}}\right\rangle_0+\rm{c.t.}\label{qo2}\\
  &=&-g\, \del_0 
\left\langle 
  T(\vec{\del_x}^2 \!\!-\!m^2)^{\frac {n-1}2}
\cO Xe^{iS_{\rm{int}}}\right\rangle_0 + g\left\langle
  T \frac{\del \cO}{\del\phi} \del_0 (\vec{\del_x}^2
      \!\!-\!m^2)^{\frac {n-1}2} \f(x)
  Xe^{iS_{\rm{int}}}\right\rangle_0+\rm{c.t.}
  \label{qo3}
\end{eqnarray}
Here there is a crucial difference to the case where we have an even
number of time derivatives in the interaction Lagrangian. We wish to
rewrite this as $\left\langle T\left(\tfrac{\delta \tilde
      S_{int}}{\delta\f(x)} X\right)\right\rangle + \rm{c.t.}$ with
the modified action
\begin{equation}
\tilde S_{int} = \int d^4x \,\cO \,\del_0(\vec{\del_x}^2
\!\!-\!m^2)^{\frac{n\!-\!1}2} \f(x) \qquad \mbox{$n$ odd}.
\end{equation}
In~\eq{qo3}, however,  one term has the time derivative acting after the time
ordering and one has it acting before the time ordering. If we wish to
write this in terms of a modified action (with all derivatives acting
after the time ordering) then we pick up an additional second order
term involving a commutator at second order in the coupling. This
extra term comes from pulling the time derivative outside the
time-ordering and has the form
\begin{equation}\ba{rcl}
  \label{eq:extra}
  &&-ig^2 \int
      d^4y\,\delta(x^0-y^0)\left\langle T\left[\frac{\del \cO}{\del\phi}
      \mbox{$(\vec{\del_x}^2   \!\!-\!m^2)^{\frac {n-1}2}$}\phi(x) ,\cO(y)  
        (\del_0)^n\f(y)\right]Xe^{iS_{\rm{int}}}\right\rangle_0\\  
&=&+g^2\,
  \left\langle
  T \left\{ (\vec{\del_x}^2 \!\!-\!m^2)^{n-1}
      (\cO \frac{\del \cO}{\del\phi})+ (\vec{\del_x}^2 \!\!-\!m^2)^{\frac
        {n-1}2} \left(\cO\frac{\del^2 \cO}{\del \phi^2} (\vec{\del_x}^2
      \!\!-\!m^2)^{\frac{n-1}2} \phi \right)\right\} 
  Xe^{iS_{\rm{int}}}\right\rangle_0 \label{disc}
\ea
\end{equation}
arising from the time derivative acting on the time ordering.

In a standard quantum field theory, 
with only a finite number of time derivatives, one can
remove this extra term using the method of finite counter
terms, and this is equivalent to using the $T*$.

Note that the above result shows that, using the Gell-Mann Low formula,
two Lagrangians which differ by total derivatives (and hence give the
same action) can nevertheless lead to
different quantum equations of motion. For example consider the case
$n=1$, $\cO=\f^m$ then the interaction Lagrangian is a total derivative
$L=\f^m\del_0\f=\del_0\f^{m+1}/(m+1)$ and so the action
 is the same as the free one. However the quantum equation of motion
 obtained using the Gell-Mann Low formula is not the same as the free
 one. In this case equation~\eq{qo3} reads
 \begin{equation}\label{ibp}
   (\square_x + m^2)\left\langle T\f(x) X\right\rangle=-g\, \del_0 
\left\langle 
  T
\phi(x)^m Xe^{iS_{\rm{int}}}\right\rangle_0 + g\left\langle
  T \del_0  \f(x)^m
  Xe^{iS_{\rm{int}}}\right\rangle_0+\rm{c.t.}
 \end{equation}

and the two terms on the right-hand-side do not cancel because one time
derivative is inside the time ordering and one outside. We obtain the
non-vanishing term~\eq{disc}.
However once again in a local theory with a finite number of time
derivatives, these discrepancies can be removed using finite counter
terms. In a theory defined with the star product however this
discrepancy may be unavoidable. Note that the fact that with the standard time
ordering Lagrangians which differ by total derivatives can lead to
different quantum theories has also been noted
in~\cite{LiaoSiboldII}. It turns out that this is not the case for the
new time ordering which we use in the next section and can thus be
seen as a further advantge of this over the standard time ordering.

This method can be extended to  more general actions. The
prescription is simple: for the quantum equation of motion, we obtain
a modified action $\tilde S_{int}$ 
simply by replacing every occurrence of $\del_t^2$ with $\vec{\del_x}^2
\!\!-\!m^2$ and leaving behind a single $\del_t$ if necessary.  

So in particular for
the $\f_*^3$ theory we {\em almost} obtain the quantum action simply
by replacing the Moyal star with the following (in momentum space)
\begin{equation}
  \label{eq:mod*}
e^{-i p\wedge q}\rightarrow \rm{cos}( p^+\wedge
 q^+)-i\,\rm{sin}( p^+\wedge q^+)\frac {p\wedge
 q}{p^+\!\!\wedge q^+\!\!}
\end{equation}
where $p^+=(\sqrt{\vec{p}^2+m^2},\vec{p})$. This is simply what one
obtains in momentum space by replacing every occurrence of
$\del_0^2$ with $\vec{\del_x}^2\!\!-\!m^2$, and leaving behind a single
$\del_0$ if you started with an odd number.  But as in the example above,
one has to be 
careful about whether the time derivatives act before or after the
time ordering. Those time derivatives which act before the time
ordering must be pulled
out of the time
ordering, leading to an additional term at order $g^2$ (as we saw
in~\eq{eq:extra}).
Furthermore in the case of non-commutative field theory this
additional term is not $SO(1,1)\xz 
SO(2)$ invariant since it involves an integral of
$\delta(x^0-y^0)[\f(x),\f^3_*(y)]$. The commutator in this case does not
give a space-like delta function needed in order to complete the
expression into a Lorentz invariant delta function as occurs for
interaction terms involving only finite numbers of time
derivatives\footnote{The breaking of the remaining symmetries in TOPT
  for non-commutative field theory was first pointed out by T.
  Reichenbach.}.

\subsection{New time ordering}

We now repeat the above calculation using the time ordering adapted to
the $SO(1,1)$ symmetries. We expect this case to preserve the
remaining symmetries for the reasons given previously.  Green's
functions are defined by the Gell-Mann Low formula~\eq{gml} together
with the time ordering defined with the $u$ coordinate as
in~\eq{eq:T2}. We wish to calculate the quantum equation of motion as
we did in the previous section for the usual time ordering. In the
case of an interaction Lagrangian containing no explicit
$u$-derivatives there will be no interference with the time ordering
and the quantum equation of motion will reproduce that of the
classical one: that is equation~\eq{line4} will be satisfied.  When
the interaction Lagrangian contains $u$-derivatives however this will
interfere with the time-ordering just as time derivatives did in the
previous section.

Consider the interaction Lagrangian
\begin{equation}
  \label{eq:sint2}
  S_{int}=g\int d^4x \,\cO\, (\del_u)^n\f(x).
\end{equation}
then using
\begin{eqnarray}
  \label{eq:ueq}
  (\square_x +m^2)\left\langle T\left( \f(x)(\del_u')^n\f(x')
\right) \right\rangle_0 &=& -i\delta(u-u')\int d^3k
(i\bar k_u)^ne^{-i\overline k(x-x')}\\
&=&-i\left( \frac{m^2-\del_2^2-\del_3^2}{2\del_v}\right)^n\delta(x-x')
\end{eqnarray}
where $d^3k=dk_vdk_2dk_3$ we find the quantum equation of motion
\begin{eqnarray}
  \label{eq:qeomu}
&&  (\square_x + m^2)\left\langle T(\f(x) X)\right\rangle \\
&=  &g(-1)^n\left(\tfrac{\del_2^2+\del_3^2-m^2}{2\del_v}\right)^n
\left\langle
  T\cO Xe^{iS_{\rm{int}}}\right\rangle_0 + g\left\langle
  T \frac{\del \cO}{\del\phi} (\del_u)^n\f(x)
  Xe^{iS_{\rm{int}}}\right\rangle_0+\rm{c.t.}\label{qeu1}\\
  &=  &g
\left\langle
  T(-1)^n\left(\tfrac{\del_2^2+\del_3^2-m^2}{2\del_v}\right)^n \cO  Xe^{iS_{\rm{int}}}\right\rangle_0 + g\left\langle
  T \frac{\del \cO}{\del\phi}
    \left(\tfrac{\del_2^2+\del_3^2-m^2}{2\del_v}\right)^n \f(y)
  Xe^{iS_{\rm{int}}}\right\rangle_0+\rm{c.t.}\label{qeu2}\\&=&\left\langle
  T\tfrac{\delta \tilde S_{int}}{\delta\f(x)} 
  X\right\rangle + \rm{c.t.}\label{qeu3}
\end{eqnarray}
where we define a modified effective action as
\begin{equation}\label{meiu}
\tilde S_{int} = g\int d^4x \cO
\left(\tfrac{\del_2^2+\del_3^2-m^2}{2\del_v}\right)^n  \f(x) \quad
\end{equation}
Note that there is here no distinction between $n$ odd and $n$ even,
and remarkably there is no complication with left over time- (ie $u$-)
derivatives 
acting both inside and outside the time ordering which were the origin
of the breaking of Lorentz invariance in the previous case (recall
that with the usual time ordering, for $n$ odd we were left with a
remaining $\del_0$ which gave extra terms and led to $SO(1,1)$
violating terms in the non-commutative case.) Here all
$u$-derivatives have disappeared and so the derivative operators can
be happily commuted through the time ordering.

Notice that to go from~\eq{qeu1} to \eq{qeu2} in the first term we
have used the fact that we can put the differential operator inside
the correlator since it commutes with the time ordering and in the
second term we have used the fact that we are dealing with free fields
in replacing $\del_u^n$.    In the case of the new
time-ordering the modified effective action is obtained by
simply replacing every occurrence of $\del_u$ by
$(\del_2^2+\del_3^2-m^2)/2\del_v$ in the action. This is because
the Gell-Mann Low formula gives time ordered vacuum expectation values
of interacting fields in terms of those for free fields for which this
replacement is possible by the free field equations of motion. 

Indeed a quicker way to deal with the complications in arriving at a
quantum equation of motion due to the time ordering is to argue as
follows. Firstly use the Gell-Mann Low
  formula to obtain an equation similar to equation~\eq{line1}. Now
  remove all $u$-derivatives using the free field equation of motion
  (since on the RHS of~\eq{line1} we have free fields.) This
  essentially involves replacing $S$ with $\tilde S$. Now simply follow
  the arguments leading to~\eq{line4} which are now valid since the
  interaction Lagrangian has no time derivatives. Clearly we end up
  with a quantum equation of motion involving the modified Lagrangian
  $\tilde S$. 

It may appear at first sight from this argument that the modified
action is the same 
as the original action and we are free to use either. It should be
noted however that the resulting modified effective action is for
vacuum expectation values of 
{\it interacting fields} (ie equations~\eqs{eq:qeomu}{qeu3} are for
interacting fields) and so $\tilde S \neq S$. Indeed we will later be
able to define Feynman rules using the modified interaction and this
is only possible once all time derivatives have been removed.

Note that in the above formulae we define the inverse differential
$1/\del_x$ via its Fourier transform as
\begin{equation}
  \label{eq:1/dv}
  \tfrac1{\del_x}f(x)=\int dk \,e^{-ikx}\frac{\tilde f(k)}{-ik}
\end{equation}
and integration by parts follows straightforwardly:
\begin{eqnarray}
  \label{eq:ibp}
  \int dx \tfrac1{\del_x}f(x) g(x)&=
  -\int dx f(x) \tfrac1{\del_x}g(x).
\end{eqnarray}
In non-commutative $\f_*^3$ field theory therefore we simply change
the Moyal star in momentum space as follows:
\begin{equation}\label{modstar}
  *=e^{-\frac i2p\wedge q} \rightarrow \overline *=e^{-\frac i2\overline p\wedge
    \overline q} \qquad
\end{equation}
where $\overline p$ is the on-shell momentum defined in~\eq{osu} to
obtain a modified action $\tilde S$, whose  na\"ive variation leads to
the quantum equations
of motion.

\subsection{The meaning of $\tilde S$}

We have shown that
\begin{equation}
\left\langle T\frac{\delta \tilde S}{\delta \f}
    X\right\rangle=\rm{c.t.} 
\end{equation} 
where $\tilde
S=\int d^4x (\del_{\mu}\phi\del^{\mu}\phi-\tfrac{m^2}{2}\phi^2)+\tilde S_{int}$
and we interpret all $u$-derivatives to be acting outside the time
ordering. This can be rewritten in terms of the generating functional
for the Green's functions $Z$ as
\begin{equation}\label{eqmZ}
 -(\square + m^2) \frac{\delta }{\delta J(x)}Z+ 
 \frac{\delta \tilde S_{int}}{\delta \f}|_{\f=\frac{\delta}{i\delta J}}Z=iJZ
\end{equation}
where the right-hand side gives the contribution of the contact terms.

The generator of connected Green's functions $Z_c$ is defined by
$Z=e^{iZ_c}$ and we define the one-point function
$\phi_c(x)=i\frac{\delta Z_c}{\delta J(x)}$. Equation~\eq{eqmZ} then
becomes (at tree level)
\begin{equation}
  (\square + m^2) \phi_c +\frac{\delta \tilde S_{int}}{\delta
  \f}|_{\f=\f_c}=J. 
\end{equation}
The above equation ignores all terms of the form
$\frac{\delta^n}{\delta J^n}Z_c$ for $n>1$. Such terms involve  at least $n-1$
closed loops and therefore vanish in the tree approximation.  The
generator of one-particle irreducible diagrams 
$\Gamma$ is then defined in the usual way as a functional of $\phi_c$:
$ \Gamma = Z_c - \int dx J \phi_c$ so that $\Gamma$ satisfies
\begin{equation}
  \frac{\delta \Gamma }{\delta \phi_c}=-J
\end{equation}
In particular, at tree level we have that
\begin{equation}
  \frac{\delta \Gamma }{\delta \phi_c}=-J=  -(\square + m^2) \phi_c
  +\frac{\delta \tilde S_{int}}{\delta  \f}|_{\f=\f_c}=\frac{\delta
  \tilde S}{\delta  \f}|_{\f=\f_c} 
  \end{equation}
  that is
\begin{equation}
\Gamma(\phi)=\tilde S(\phi)  
\end{equation}
or in other words $\tilde S$ is the tree level effective action.

This is a somewhat remarkable result: usually the zero loop approximation
to $\Gamma$ can be identified with the classical action. And the vertices of
the classical action are used as the vertices in the interaction as
defined, say via the Gell-Mann Low formula. Here however the zero-loop approximation
to the vertex functional $\Gamma$ cannot be identified with the classical
action but differs by the transition to the mass shell within the star
product vertices as enforced by the quantization procedure.
It is also to be noted that this on-shell star product is not really a star
product: it is e.g.\ not associative.

\section{Symmetries}

We wish to prove explicitly that the theory defined via the Gell-Mann Low formula
and with the modified time ordering is invariant under translations
and the remaining $SO(1,1)\xz SO(2)$ symmetry (at least at tree
level).

For this we simply have to show that the effective action $\Gamma$
is invariant under these symmetries.
  If it is, then we will also be able to construct
conserved energy-
and angular-momentum tensors by Noether's theorem.

\subsection{Translations: the energy momentum tensor}

Since the effective action $\Gamma$ has no explicit $x$ dependence, it
must be translation invariant.  We do not consider here the free part
of the effective action which takes the standard form and gives the
standard energy momentum tensor.  The interaction part has the form
\begin{equation}
  \Gamma_{int}=\int dx\phi_1\overline *\phi_2\overline *\phi_3= \int d\mu(p_i) \ \tilde
    \phi(p_1) \tilde \phi(p_2) 
    \tilde\phi(p_3) F(\overline p_1,\overline p_2,\overline p_3)
\end{equation}
with $d\mu(p_i):= dp_1dp_2dp_3
  \delta(p_1+p_2+p_3)$ and where $\tilde\phi$ is the Fourier transform
  of $\phi$ and $F$ is the non-commutative phase factor 
\begin{equation}
  F(p_1,p_2,p_3)=e^{-i\left( p_1\wedge p_2 +
        p_1\wedge p_3 +p_2\wedge
        p_3\right)}.
\end{equation}
Recall that $\overline *$ is defined (in momentum space)
in~\eq{modstar} and $\overline p$ is the on-shell momentum defined in~\eq{osu}.
Explicitly, the Ward identity for infinitesimal translations has the
form
\begin{eqnarray}\label{dG}
  \delta \Gamma &=& \int dx  (a^{\mu} \del_{\mu} \phi)\overline*\phi\overline*\phi+
    \phi\overline*(a^{\mu} \del_{\mu}
    \phi)\overline*\phi+\phi\overline*\phi\overline*(a^{\mu} \del_{\mu} 
    \phi)\\
  &=& \int d\mu(q,p_i)\ 
    \tilde \phi(p_1) \tilde \phi(p_2)
    \tilde\phi(p_3) \tilde a^{\mu}(q)(-i)I_{\mu}(q,p_1,p_2,p_3)\label{Im}
\end{eqnarray}  
with $d\mu(q,p_i):=dq dp_1 dp_2 dp_3 \delta(q+p_1+p_2+p_3)$
and where
\begin{eqnarray}
  I_{\mu}(q,p_1,p_2,p_3)&=&
    p_{1\mu}
      F(\overline{q+p}_1,\overline p_2,\overline p_3)+p_{2\mu}F(\overline p_1,\overline
      {q+p_2},\overline p_3)+p_{3\mu}F(\overline p_1,\overline p_2,\overline {p_3+q})\\
      &=& 
    (p_{1}+p_{2}+p_{3})_{\mu}\ 
     F(\overline p_1,\overline p_2,\overline p_3)+
     \ba[t]{l}
     p_{1\mu}(F(\overline {q+p_1},\overline p_2,\overline p_3)-F(\overline p_1,\overline p_2,\overline p_3))\\
     p_{2\mu}(F(\overline p_1,\overline {q+p_2},\overline p_3)-F(\overline p_1,\overline p_2,\overline p_3))\\
     p_{3\mu}(F(\overline p_1,\overline p_2,\overline {q+p_3})-F(\overline p_1,\overline p_2,\overline p_3)).
     \ea
\end{eqnarray}

Now
\begin{eqnarray}
  F(\overline {p_1+q},\overline p_2,\overline p_3)-F(\overline p_1,\overline p_2,\overline p_3)
  &\sim&F(\overline p_1,\overline p_2,\overline p_3)\xz\Phi_1\xz(-i\,q_1\wedge(\overline p_{2}+\overline p_{3}))\label{FF1}\\
  F(\overline p_1,\overline {p_2+q},\overline p_3)-F(\overline p_1,\overline p_2,\overline p_3)
  &\sim&F(\overline p_1,\overline p_2,\overline p_3)\xz\Phi_2\xz(-i\,q_2\wedge(\overline p_{3}-\overline p_{1}))\label{FF2}\\
  F(\overline p_1,\overline p_2,\overline {p_3+q})-F(\overline p_1,\overline p_2,\overline p_3)
  &\sim&F(\overline p_1,\overline p_2,\overline p_3)\xz\Phi_3\xz(i\,q_3\wedge(\overline p_{1}+\overline p_{2}))\label{FF3}
\end{eqnarray}
where
\begin{eqnarray}
  \Phi_1&:=& \left( \frac{e^{i(\overline {p_2+p_3}
      +\overline p_{1}) \wedge(\overline p_{2}+\overline p_{3})}-1}{i(\overline {p_2+p_3}
    +\overline p_{1})\wedge(\overline p_{2}+\overline p_{3})}\right)\\
\Phi_2&:=& \left( \frac{e^{i(\overline {p_1+ p_3}
      +\overline p_{2}) \wedge(\overline p_{3}-\overline p_{1})}-1}{i(\overline {p_1+p_3}
    +\overline p_{2})\wedge(\overline p_{3}-\overline p_{1})}\right)\\
\Phi_3&:=& \left( \frac{e^{-i(\overline p_1+\overline {p_2
      +p_{3}}) \wedge(\overline p_{1}+\overline
      p_{2})}-1}{-i(\overline p_1+\overline {p_2
    +p_{3}})\wedge(\overline p_{1}+\overline p_{2})}\right).
\end{eqnarray}
We have here defined 
\begin{equation}
  q_i:=\overline {q+p_i} -\overline p_{i}
\end{equation}
so three of the components of $q_i$ are equal to those of $q$ ie
$(q_i)_v=q_v,\ (q_i)_2=q_2,\ (q_i)_3=q_3$ whereas the $u$-th component
is for example
\begin{equation}
  (q_1)_u= \left(
      \frac{m^2+(q+p_1)_a(q+p_1)_a}{2(q_v+p_{1v})}-\frac{m^2+p_ap_a}{2p_{1v}} 
      \right)
      \sim 
      \frac{q_v(m^2+p_{1a}p_{1a})-q_a(p_1-p_2-p_3)_ap_{1v}}{2(p_{2v}+p_{3v})p_{1v}}.
  \end{equation}

The `$\sim$' in all the above equations means `equal when multiplied by
$\delta(q+p_1+p_2+p_3)$': 
we have used the delta-function to
ensure that we have an expression which is linear in $q$
(corresponding to a single derivative of $a_{\mu}$.)
Now define $S_{i\mu}$ via 
\begin{eqnarray}
  q_1\wedge(\overline p_{2}+\overline p_{3})&=&q^{\mu}S_{1\mu}\\
  q_2\wedge(\overline p_{3}-\overline p_{1})&=&q^{\mu}S_{2\mu}\\
  -q_3\wedge(\overline p_{1}+\overline p_{2})&=&q^{\mu}S_{3\mu}.
\end{eqnarray}
In this way we have obtained an expression for $I_{\mu}$ which is linear in
$q$ 
\begin{equation}
  I_{\mu}\sim F(\overline p_1,\overline p_2,\overline p_3)\left( -q_{\mu}
    -i\,q_{\nu} \sum_i S_{i}^{\nu} p_{i\mu} \Phi_i \right)
\end{equation}
Putting this into~\eq{Im} gives
\begin{equation}
  \delta\Gamma=-\int dx \del_{\nu}a^{\mu}(x)T_{\mu}^{\nu}
\end{equation}
where
\begin{equation}
  (T_{int})_{\mu}^{\nu} =\delta_{\mu}^{\nu} \Gamma_{int} + i\int dp_1 dp_2
  dp_3 e^{-ix(p_1+p_2+p_3)}
  \tilde \phi(p_1) \tilde \phi(p_2)
  \tilde\phi(p_3)F(\overline p_1,\overline p_2,\overline p_3)\sum_i
  S_{i}^{\nu}p_{i\mu}\Phi_i 
\end{equation}
for $a,b\in\{2,3\}$.

\subsection{Lorentz transformations: angular momentum tensor}

The effective action is also invariant under the remaining $SO(1,1)\xz
SO(2)$ transformation. 
The underlying reason why the effective action is invariant under these
symmetries is that the symmetries commute with the projection of $p$
onto the mass-shell $p\rightarrow \overline p$. In other words we have
$\overline {p+\delta p}=\overline p +\delta \overline p$ where $\delta$ is an infinitesimal
$SO(1,1)\times SO(2)$ transformation. We show this explicitly in
equation~\eq{dp}

Note that there is another crucial difference with the standard time
ordering here. With the standard time ordering
one projects onto the mass-shell by replacing $p_0$ with $\pm
\sqrt{p_1^2+p_{a}p^{a}+m^2}$  instead of replacing $p_u$ as we do
with the new time ordering. This projection does not commute with
the $SO(1,1)\times SO(2)$ transformation thus leading to a loss of the
symmetry.

The explicit proof of covariance of the effective action and
construction of the angular-momentum tensor follows in a similar way to
that of the energy-momentum tensor.
An infinitesimal Lorentz transformation has the form~\eq{dG}
with $a^{\mu}=w^{\mu}_{\nu} x^{\nu}$ and so $\delta \Gamma$ can be
written 
\begin{equation}
  \delta \Gamma =\int d\mu(p_i,q)\ 
 \tilde w^{\mu\nu}(q)
  \left(p_{1\nu}\del_{\mu}\tilde \phi(p_1) \tilde \phi(p_2)
    \tilde\phi(p_3) F(\overline {q+p_1},\overline p_2,\overline p_3) + \dots
  \right)
\end{equation}
where the dots indicate two more similar terms.
We proceed as for the energy-momentum tensor we write $F(\overline
{q+p_1},\overline p_2,\overline p_3)$ as
$F(\overline p_1,\overline p_2,\overline p_3)+(F(\overline {q+p_1},\overline p_2,\overline p_3)- F(\overline p_1,\overline p_2,\overline p_3))$ and similarly for
the other two terms. Using~\eqs{FF1}{FF3} and integration by parts in momentum
space we arrive at
\begin{eqnarray}\label{inv}
  \delta \Gamma &=& -\int dx\, \del_{\nu} w^{\mu\nu} x_{\mu}
  (\phi\overline *\phi\overline *\phi) - w^{\mu\nu} \int d \mu(p_i)\ \sum_i
  p_{i\mu}\tfrac{\del}{\del p_{i\nu}} F(\overline p_1,\overline p_2,\overline p_3) \\
  &&+ \int dx \del_{\rho}w^{\mu\nu} \int e^{-ix(p_1+p_2+p_3)}
  \tilde \phi(p_1) \tilde \phi(p_2)
  \tilde\phi(p_3) F(\overline p_1,\overline p_2,\overline p_3)
  \nonumber \\
  &&\qquad \qquad \qquad \qquad \times \sum_i
  S_{i}^{\rho} \Phi_i  p_{i\mu} \tfrac{\del}{\del p_{i\nu}} (
  \tilde \phi(p_1) \tilde \phi(p_2) \tilde\phi(p_3) ) \nonumber
\end{eqnarray}
Now the first and third terms vanish for a global Lorentz
transformation (for which  $w^{\mu\nu}$ is constant).
The second term represents an infinitesimal Lorentz transformation in
momentum space of $F(\overline p_1,\overline p_2,\overline p_3)$. Note that 
\begin{eqnarray} \label{dF}
\delta F(\overline p_1,\overline p_2,\overline p_3)&:=&  \sum_i w^{\mu}{}_{\nu}
  p_{i\mu}\tfrac{\del}{\del p_{i\nu}}
  F(\overline p_1,\overline p_2,\overline p_3)\\
  &=&F(\overline p_1,\overline p_2,\overline p_3) \sum_i w^{\mu}{}_{\nu}
  p_{i\mu}\tfrac{\del}{\del
  p_{i\nu}}  (\overline p_{1}\wedge \overline p_{2}+\overline p_{2}\wedge \overline p_{3}+\overline p_{1}\wedge \overline p_{3})
\end{eqnarray}
Now for an infinitesimal $SO(1,1)\times SO(2)$ transformation 
$  w^u{}_u=-w^v{}_v$ and $w^2_3=-w^3_2$ are the only non-zero
components of $w^{\mu}_{\nu}$ and one can  easily show that
\begin{equation}\label{dp}
  w^{\mu}{}_{\nu} p_{\mu}\del^{\nu}\overline p_{\rho}=\overline p_{\mu}w^{\mu}{}_{\rho}
  \qquad w \in so(1,1)\times so(2)
\end{equation}
which is the statement that $\overline p$ is covariant under $SO(1,1)\times
SO(2)$. Note that this is not true  for an arbitrary
infinitesimal Lorentz 
transformation. In particular there are extra non-covariant terms in
the above equation involving $w^u{}_{a}$.  

It is now easy to see that $\delta F=0$ under an $SO(1,1)\times SO(2)$
transformation since we know that $\theta^{\mu\nu}$ is
invariant (ie $w^{\mu}{}_{\mu'}\theta^{\mu'\nu}+
\theta^{\mu\nu'}w^{\nu}{}_{\nu'}=0$).

Having thus proven invariance of the tree-level quantum theory under
$SO(1,1)\times SO(2)$ we can then construct the angular-momentum tensor
simply by reading  off the coefficient of $\del_{\rho}w^{\mu \nu}$ in~\eq{inv}.
We obtain 
\begin{equation}
M^{\rho\mu\nu}=  \eta^{\rho [\nu}x^{\mu]}\Gamma_{int}+ \int dp_1 dp_2
  dp_3 e^{-ix(p_1+p_2+p_3)} F(\overline p_1,\overline p_2,\overline p_3)
  \sum_i
  S_{i}^{\rho}  \Phi_i p_{i}^{[\mu}\del^{\nu]} \left(\tilde \phi(p_1)
  \tilde \phi(p_2) \tilde\phi(p_3)\right) 
\end{equation}

\section{LSZ reduction}
In ordinary quantum field theory one obtains matrix elements of operators
from time ordered Green functions by the LSZ reduction formulae. Since they
are based on the usual time ordering with respect to $x^0$ we have to check
that also in our case we have analogous relations.\\
The quantity most immediately associated to Green functions of interacting
fields is the $S$-matrix. One can straightforwardly mimic the
manipulations used to derive  the standard LSZ reduction formulae if
we postulate the existence of an asymptotic (weak) limit 
\begin{equation}
\sqrt{z} \Phi_{\rm{in}}(x)=\lim_{u\to -\infty} \Phi(x)
 \qquad
\sqrt{z} \Phi_{\rm{out}}(x)=\lim_{u\to +\infty} \Phi(x)
\end{equation}
Here the factor $z$ corresponds to the wave function renormalisation and will be
suppressed in the formulae to follow.
The result is that 
an arbitrary matrix element with $n$ out- and $l$ incoming particles
is related to Green functions of interacting fields as follows:
\begin{eqnarray}
  &&{}_{\rm{out}}\langle p_1 \cdots p_n| q_1
  \cdots q_l\rangle_{\rm{in}}= {}_{\rm{in}}\langle p_1 \cdots p_n|S| q_1
  \cdots p_l \rangle_{\rm{in}} \\
  &&= i^{n+l} \int d^4y_1 \cdots d^4x_l e^{i p_ky_k
  +q_jx_j} (\square_{y_1}+m^2) \cdots (\square_{x_l}+m^2) \langle 0
  |T\Phi(y_1) \cdots \Phi(x_l) |0\rangle\\
  &&=(-i)^{n+1} (p_1^2-m^2)\cdots (q_l^2-m^2) \tilde G(p_1,\cdots,q_l)|
\end{eqnarray}
where $\tilde G$ is the Fourier transform of the time ordered Green's
function and the vertical line indicates that all momenta are put
on-shell by setting $p_u=(p_{a}p_{a}+m^2)/(2p_v)$, $S$ is the
$S$-matrix and the time ordering is with respect to the $u$
component.

\section{Comparison of Feynman rules for different formulations of
  non-commutative field theories}

\subsection{The na\"ive Feynman rules}\label{naivef}

These correspond to taking the Gell-Mann Low formula but assuming that
all derivatives in the interaction Lagrangian occur outside the
time-ordering.
We sketch the na\"ive momentum space Feynman rules (up to factors) for
 $\f_*^3$ theory 
looking at a diagram with $N$ internal lines, with momenta $k_i$,
 $E$ external lines with momentum $p_i$ and $V$ 
 vertices. 
 The $j$th vertex has lines entering it with momenta $r_j,s_j,t_j$
 which could be internal or external lines.
 The Feynman rules for this diagram as given for example by
 Filk~\cite{Filk:1996dm} result in 
\begin{eqnarray}\label{nF}
  \tilde G(p_i)&\sim& S^{-1}\prod_{i=1}^E P(p_i)\prod_{i=1}^N \int
  {d^4k_i}  P(k_i) \times
  \prod_{j=1}^V\delta^4(r_{j}+s_{j}+t_{j}) \times
  F(r_{j},s_{j},t_{j})
\end{eqnarray}
where $S$ is a symmetry factor, $P(k_i)=\frac
i{k_i^2-m^2+i\epsilon}$ is the Fourier 
transform of the propagator and $F(p_1,p_2,p_3)$ is the non-commutative
phase factor at the vertex given by 
$\f_*^3$ theory. 
We have
\begin{eqnarray}
F(p_1,p_2,p_3)&=&\sum_{\sigma \in
  P_3}e^{i(p_{\sigma(1)}\wedge p_{\sigma(2)}+p_{\sigma(1)}\wedge p_{\sigma(3)}+p_{\sigma(2)}\wedge p_{\sigma(3)})}
\end{eqnarray}
where $P_3$ is the set of permutations of $(1,2,3)$.
The momenta $r_j,s_j,t_j$ are the momenta entering a vertex $V_j$ and
can be read off from the Feynman diagram. We illustrate with a
1-loop two-point function below.

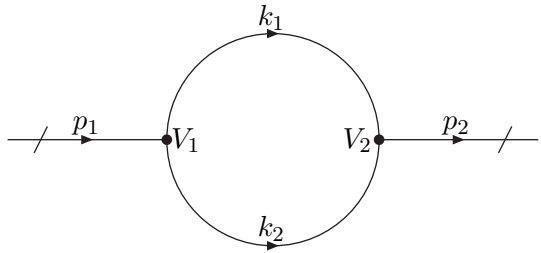
{\begin{center}
\begin{picture}(300,100)(0,0)
\ArrowLine(50,50)(110,50)\Text(80,52)[b]{$p_1$} 
\Line(60,45)(65,55) 
\Vertex(110,50){2}\Text(112,50)[l]{$V_1$}
\ArrowArc(150,50)(40,180,0)\Text(150,92)[b]{$k_1$}
\ArrowArcn(150,50)(40,180,360)\Text(150,12)[b]{$k_2$}
\ArrowLine(190,50)(250,50)\Text(220,52)[b]{$p_2$} 
\Line(235,45)(240,55)
\Vertex(190,50){2}\Text(188,50)[r]{$V_2$}
\end{picture}
\\
{\sl 1-loop diagram. The momenta $r_j,s_j,t_j$ can be read off from the
  diagram: $(r_1,s_1,t_1)=(p_1,-k_1,-k_2) \quad (r_2,s_2,t_2)=(k_1,k_2,-p_2)$}
\end{center}
}

So the Green's function corresponding to this diagram is
\begin{eqnarray}\nonumber
  \tilde G(p_1,p_2)&=&S^{-1} P(p_1)P(p_2)\int
  {d^4k_1}{d^4k_2}  P(k_1)P(k_2) 
  \delta^4(p_1-k_1-k_2)\delta^4(k_1+k_2-p_2)\\
  &&\times F(p_1,-k_1,-k_2)F(k_1,k_2,-p_2)
  \end{eqnarray}
It is by now well-known that these rules respect unitarity only in the case that
$\theta_e$ vanishes since otherwise there is a conflict of commuting time derivatives
of the star product with time ordering.

\subsection{TOPT time ordered with respect to $x^0$.}\label{TOPT}

In~\cite{LiaoSiboldI} Feynman rules were also derived  by following the Gell-Mann Low 
formula using the usual (ie with respect to $x^0$) time ordering but proper care had been taken
to the occurrence of time derivatives from the star product before
time ordering. These rules also follow from the Hamiltonian approach
of~\cite{DoplicherFredenhagenRoberts}.
We now associate
a number $\lambda_{i}=\pm 1$ with each internal momentum $k_i$ and a
number $\mu_{i}=\pm 1$ with each external momentum $p_i$.
The
resulting Feynman rules are
 \begin{eqnarray}
  \tilde G(p_i)&\sim&
  S^{-1}\sum_{\lambda_i,\mu_i}\prod_{i=1}^E
  P_{\mu_i}(p_i)
  \prod_{i=1}^N \int 
  {d^4k_i}  P_{\lambda_{i}}(k_i)
  \prod_{j=1}^V\delta^4(r_{j}+
  s_j+t_j)
  F(r^{\lambda}_j,s^{\lambda}_j,t^{\lambda}_j)
  .\label{TF}
\end{eqnarray}
Here the momenta appearing in the phase factor are put on-shell by
replacing the zeroth component of $p$, with $\lambda w_{p}$ as
indicated by the superscript $\lambda$. The notation here is somewhat 
schematic: the superscript $\lambda$ is that associated with the
momentum $r_j,s_j $or $t_j$. 
We must then sum over $\lambda_i=\pm 1$ corresponding
to positive and
negative frequency momenta. The factor 
\begin{eqnarray}
  P_{\lambda}(k)&=&\frac{\lambda}{2w_k(k^0-\lambda(w_k-i\epsilon))} =
  \frac{\eta_{\lambda}(k)}{k^2-m^2+i\epsilon}\\ 
  \eta_{\lambda}(k)&=&1/2(1+\lambda k_0/w_k)\\
  w_k&=& \sqrt{\vec k^2+m^2}
\end{eqnarray}
is the Fourier transform of $\theta(\lambda x^0) D^\lambda(x)$. Note
that for on-shell momenta this is equal to the propagator
whereas even for off-shell momenta we have 
that $P_+(k)+P_-(k)=P(k)$. This implies that if the non-commutative phase
factor is independent of $\lambda$ (as for example in the case of pure
space-space non-commutativity) then summing over $\lambda$ we obtain
the na\"ive Feynman rules~\eq{nF}. Since, however, the phase factor
does explicitly
depend on $\lambda$ in the generic case (ie with $\theta_e\neq 0$) we can not
re-express these rules in terms of ordinary  propagators.

For the 1-loop diagram above we obtain 
\begin{equation} \ba{rcl}
  \tilde G(p_1,p_2)&=&S^{-1}
  \sum_{\lambda_i,\mu_i}P_{\mu_1}(p_1)P_{\mu_2}(p_2)\!\!\int \!\!
  {d^4k_1}{d^4k_2}  P_{\lambda_1}(k_1)P_{\lambda_2}(k_2) 
  \mbox{\small $\delta(p_1\!-\!k_1\!-\!k_2)\delta(k_1\!+\!k_2\!-\!p_2)$}\\
  &&\times F(p_{1\mu_1},-k_{1\lambda_1},-k_{2\lambda_2})
  F(k_{1\lambda_1},k_{2\lambda_2},-p_{2\mu_2})  
\ea
\end{equation}

These rules lead to unitary amplitudes also in the case when $\theta_e$
does not vanish, but still they are not satisfactory: the main drawback
being that the underlying $SO(1,1)\xz SO(2)$-invariance is not maintained.

\subsection{TOPT with the new time-ordering}

Finally we consider the Feynman rules obtained from the
Gell-Mann Low formula with the new time ordering introduced in
section~\ref{lwnto}. We could derive the Feynman rules from first
principles, but in order to compare this approach with the previous
one, we instead 
derive the new Feynman rules by suitably adapting~\eq{TF}. 
We  only need to find modifications  
for $P_{\lambda}(k)$ and for the non-commutative phase factor.
Since $P_{\lambda}(k)$ is the Fourier transform of $\theta(\lambda
x^0) D^\lambda(x)$ we replace this with the Fourier transform of
$\theta(\lambda u) D^\lambda(x)$.
Using
\begin{eqnarray}
  \theta(\lambda u)&=&\frac{i \lambda}{2\pi} \int \frac{ds e^{-is
      u}}{s+i\epsilon\lambda}\\
  D^{\lambda}(x)&=&\int \frac{d^3 p}{2p_v} \theta(\lambda
      p_v)e^{-i\overline p x}
\end{eqnarray}
 we obtain 
\begin{equation}
  P_{\lambda}(k) \rightarrow \frac{\theta(\lambda k_v)}{2k_v
  (k_u-\bar k_u+i\epsilon\lambda)}=\frac{\theta(\lambda k_v)}{k^2-m^2+i\epsilon}.
\end{equation}

The non-commutative phase factor is obtained by taking the na\"ive
phase factor $F$, putting all the momenta on-shell,  and splitting
into positive and negative frequency parts. In the present case we put
the momenta on-shell by replacing $p$ with $\overline p$ (see~\eq{osu}) and the
positive and negative frequency parts correspond to positive and
negative $p_v$. So we expect the non-commutative phase factor
\begin{eqnarray}
   F(k^{\lambda_{1}}_{1},k^{\lambda_{2}}_{2},
  k^{\lambda_{3}}_{3}) \rightarrow 
  \theta(\lambda_1k_{1v})\theta(\lambda_2k_{2v})\theta(\lambda_3k_{3v})
  F(\overline k_1,\overline k_2,\overline k_3).
\end{eqnarray}
Note that in this case, splitting into positive and negative frequency
parts simply corresponds to taking $p_v$ positive or negative (which
we have indicated by using step functions). However the step functions
are already present in $P_{\lambda}(x)$ so the phase factor is
effectively independent of $\lambda$. Indeed, if we perform the sum over
$\lambda$ the $P_{\lambda}$s  sum to give complete propagators and we
obtain  the following Feynman rules
\begin{eqnarray}
  \tilde G(p_i)&\sim& S^{-1}\prod_{i=1}^E P(p_i)\prod_{i=1}^N \int
  {dk_i}  P(k_i) \times
  \prod_{j=1}^V\delta(r_j+s_{j}+t_{j})
  \times F(\overline r_j,\overline t_j,\overline s_j).
\end{eqnarray}
Note that the only difference to the na\"ive case is the appearance of
the modified phase factor.

So the Green's function corresponding to the 1-loop diagram above is
\begin{eqnarray}\nonumber
  \tilde G(p_1,p_2)&=&S^{-1} P(p_1)P(p_2)\int
  {d^4k_1}{d^4k_2}  P(k_1)P(k_2) 
  \delta^4(p_1-k_1-k_2)\delta^4(k_1+k_2-p_2)\\
  &&F(\overline p_1,-\overline k_1,-\overline k_2)F(\overline k_1,\overline k_2,-\overline p_2)\label{1lu}
  \end{eqnarray}
  Clearly this is a tremendous simplification when actually
  calculating diagrams.  In a sense the new time ordering rendered
  superfluous the explicit distinction between positive and negative
  frequency parts and thus reunited what had to be separated in old
  fashioned time ordered perturbation theory.  It also seems to obey
  the positive energy condition discussed in~\cite{fujikawa} since the
  free propagator certainly does and in diagrams describing
  interaction also energy components occur with the correct signs
  only. This is ensured by the fact that we can formulate Feynman rules in
  terms of propagators.
  
  Of course, as for TOPT~\cite{LiaoSiboldII} and the equivalent
  Hamiltonian formulation~\cite{DoplicherFredenhagenRoberts} unitarity
  is now automatic due to a correct treatment of the time ordering. In
  the appendix we show this explicitly by checking the
  optical theorem.

\label{UTO}

\section{Discussion, conclusions and outlook}

The starting point for the considerations of the present paper is the
symmetry content of a theory of quantum fields if one has interactions
according to the Moyal product and a generic $\theta_{\mu\nu}$. It
comprises translations and $SO(1,1)\xz SO(2)$ which should be
maintained in the course of quantization. Basing time ordering on the
values $x^0$ of the coordinates this is not the case. It is however
true when we time order according to the light wedge variable $u=(x^0
- x^1)/\sqrt{2}$. We proved the symmetry content of the theory to be
the desired one by explicitly constructing the respective conserved
currents in the form of Ward identities for time ordered Green
functions. Here we used the gratifying fact that the quantum equations
of motion can be written in closed form via an effective action. It is
remarkable that this effective action is in the tree approximation not
the classical action but the classical action with star product
modified into products living on the mass
shell of the fields.\\
A further noticeable simplification arises on the level of Feynman
rules. Again as a consequence of the new time ordering we arrive at
essentially na\"ive ones: propagators are the usual ones, phase
factors are those of the modified star
product ie mass shell factors written in the new variables.
Note that one might be worried about infrared divergences
arising from using light cone coordinates (as for example
$p_v\rightarrow 0$). These Feynman rules show that these are unlikely
to occur since the only divergent pieces occur
in the phase where they are rendered harmless.\\
Unitarity has been checked to hold in an explicit example which
however permits immediate generalisation on a formal level. Hence the
theory is certainly
well defined on the tree level. \\
LSZ reduction works well when the limit of $u$ going to plus or minus
infinity is taken as defining the asymptotics. Causality is lost in
the sense that there are in general no two points $x,y$ in space-time
where we can be certain that $\Phi(x)$ commutes with $\Phi(y)$ (ie no
analogue of `space-like separation').  This means that our time
ordering defines a genuine `before' and `after'. There is no ambiguous
(space-like) region as there is in an ordinary relativistic quantum
field theory.
\\
Thinking of extensions of our results one can indeed have the hope
that gauge theories exist as well for generic $\theta$ since global
symmetry currents will exist due to the simple form of the quantum
equations of motion. Hence the examples which are known to exist for
vanishing $\theta_e$ should all be generalisable to generic $\theta$.
In the actual formulation of gauge fixing and BRS invariance the
expertise collected in light cone quantization should be helpful. For
higher orders analogously one should at least be able to construct
what can be constructed for restricted $\theta$. Since with the time
ordering the integrals truly change one should also have a fresh look
at the ultraviolet/infrared connection. It may very well
differ from the previous one.\\

{\bf Acknowledgements}

We would like to thank Yi Liao, Christoph Dehne and Tobias Reichenbach
for helpful discussions and in particular we thank Tobias Reichenbach
for making results from his diploma thesis available prior to
publication. We would also like to thank the referee for many helpful
questions and coments.

\section*{Appendix:Unitarity}

Having derived the Feynman rules we are now in a position to test
unitarity of the theory by checking the optical theorem. As an example
we study the two point function of $\f_*^3$ in the one loop approximation.
The optical theorem can be given diagrammatically as

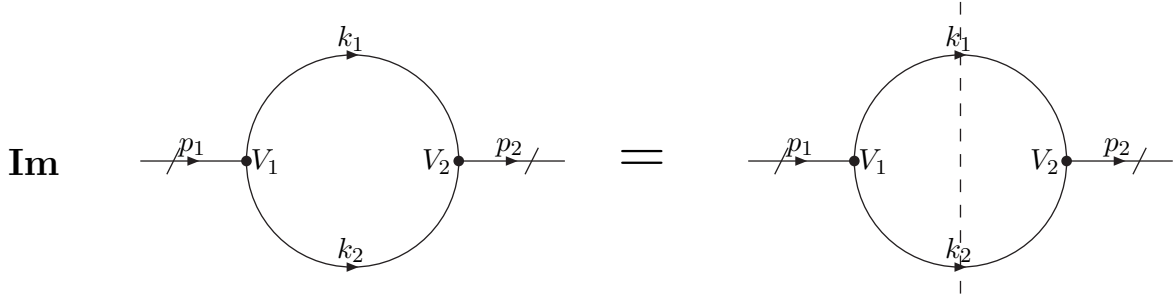
\begin{center}

\begin{picture}(450,100)(0,0)

\Text(0,50)[]{\Large \bf Im}

\SetOffset(-30,0)
\ArrowLine(70,50)(110,50)\Text(90,52)[b]{$p_1$} 
\Line(80,45)(85,55) 
\Vertex(110,50){2}\Text(112,50)[l]{$V_1$}
\ArrowArc(150,50)(40,180,0)\Text(150,92)[b]{$k_1$}
\ArrowArcn(150,50)(40,180,360)\Text(150,12)[b]{$k_2$}
\ArrowLine(190,50)(230,50)\Text(210,52)[b]{$p_2$} 
\Line(215,45)(220,55)
\Vertex(190,50){2}\Text(188,50)[r]{$V_2$}

\Text(260,50)[]{\Huge $=$}

\SetOffset(200,0)
\ArrowLine(70,50)(110,50)\Text(90,52)[b]{$p_1$} 
\Line(80,45)(85,55) 
\Vertex(110,50){2}\Text(112,50)[l]{$V_1$}
\ArrowArc(150,50)(40,180,0)\Text(150,92)[b]{$k_1$}
\ArrowArcn(150,50)(40,180,360)\Text(150,12)[b]{$k_2$}
\ArrowLine(190,50)(230,50)\Text(210,52)[b]{$p_2$} 
\Line(215,45)(220,55)
\Vertex(190,50){2}\Text(188,50)[r]{$V_2$}
\DashLine(150,110)(150,0){5}
\end{picture}
\\
{\sl Optical theorem for the  1-loop diagram. The imaginary part of the
amputated 1-loop diagram equals the cut graph on the right which will
be defined below.}
\end{center}

The proof of unitarity follows closely that of~\cite{LiaoSiboldI} for the usual
time-ordering.
The left hand side of this equation is the imaginary part of the 1
loop function given in~\eq{1lu}, amputated by removing the terms
$P(p_1)P(p_2)$.
 In order to find the imaginary part of this we first explicitly perform 
the integration over the $u$th component of all internal momenta
$(k_i)_u$. For this it is crucial that unlike the na\"ive
Feynman rules, here  the non-commutative phase factor
$F$ is independent of $(k_i)_u$. The integration over $(k_1)_u$
can be finished by the delta function and  the integration over $k_2$ can
be performed using contour
integration.
The result is 
\begin{eqnarray}\nonumber
  \frac{\tilde G(p_1,p_2)}{P(p_1)P(p_2)}&=&S^{-1} \int
  {d^3\mu_1}{d^3\mu_2} 
  \frac{\delta^3(p_1-k_1-k_2)\delta^4(p_1-p_2)}{p_{1u}-\overline
  k_{1u}-\overline k_{2u}+i\epsilon}\\
  &&\times F(\overline p_1,-\overline k_1,-\overline k_2)F(\overline k_1,\overline k_2,-\overline
  p_2)
\end{eqnarray}
where $d^3\mu_i=dk_vdk_2dk_3/2k_v$ is the invariant measure.

We are now in a position to compute the imaginary part of this. Using the
distribution identity
\begin{eqnarray}
  Im\left(\frac 1{x+i\epsilon}\right)=\delta(x+i\epsilon)
\end{eqnarray}
we find the left hand side of the optical theorem
\begin{eqnarray}\nonumber
Im\frac{\tilde G(p_1,p_2)}{P(p_1)P(p_2)}&=&S^{-1} \int
  {d^3\mu_1}{d^3\mu_2} 
  \delta^4(p_1-\overline k_1-\overline k_2)\delta^4(p_1-p_2)
   F(\overline p_1,-\overline k_1,-\overline k_2)F(\overline k_1,\overline k_2,-\overline
  p_2)\\\nonumber
  &=&S^{-1} \int
  {d^4k_1}{d^4k_2} 
  \delta^4(p_1- k_1-k_2)\delta^4(k_1+k_2-p_2)
   \delta^4(k_1^2-m^2)\delta^4(k_2^2-m^2)\\
   &&\times F(\overline p_1,-\overline k_1,-\overline k_2)F(\overline k_1,\overline k_2,-\overline
  p_2).
\end{eqnarray}
Comparing with~\eq{1lu} we see that the imaginary part of the
amputated Green's function is obtained simply by replacing the propagators
with delta functions $P(k)\rightarrow \delta(k^2-m^2)$. 

The
right-hand side of the optical theorem is defined to be
\begin{eqnarray}
  \int d^4k_1d^4k_2
  \delta(k_1^2-m^2)\delta(k_2^2-m^2){\cal M}^*(-p_2\rightarrow
  k_1k_2){\cal M}(p_1\rightarrow k_1k_2)
\end{eqnarray}
and by substituting in 
\begin{eqnarray}
  {\cal M}(p_1\rightarrow k_1k_2)&=&F(\overline
  p_1,-k_1,-k_2)\delta^4(p_1-k_1-k_2)\\
{\cal M^*}(p_2\rightarrow k_1k_2)&=&F(\overline
  p_1,-k_1,-k_2)\delta^4(p_1-k_1-k_2)
\end{eqnarray}
we find that the right-hand side equals the left-hand side and the 
optical theorem is satisfied.\\
As usual, in this unitarity check one had to be sure only of the fact that the
imaginary part of the loop diagram is finite. The real part diverges and would
require proper definition which we do not attempt here. On this formal level
one can also state generalisations: unitarity will be alright to all orders 
with our time ordering since our noncommutative phase factors do not change the
unitarity character of an underlying unitary theory.

 
\providecommand{\href}[2]{#2}\begingroup\raggedright\endgroup

\end{document}